\documentclass[10pt, journal]{IEEEtran}
\usepackage{cite}
\usepackage{amsmath,amssymb,amsfonts}
\usepackage[noend]{algorithmic}
\usepackage{graphicx}
\usepackage{textcomp}
\usepackage{xcolor}
\usepackage{amssymb}
\usepackage{indentfirst}
\usepackage{amsmath}
\usepackage{bm}
\usepackage{booktabs}
\usepackage{epsfig}
\usepackage{algorithm}
\usepackage{changepage}
\usepackage{amsthm} 
\usepackage{soul,color}
\usepackage{hyperref}
\usepackage[font=small]{caption}
\usepackage{subcaption}
\usepackage{soul}

\usepackage{tikz}
\usepackage{pgfplots}
\usetikzlibrary{plotmarks}
\usetikzlibrary{matrix,positioning,shapes,arrows,decorations,calc,fit}
\usetikzlibrary{decorations.pathreplacing}
\usetikzlibrary{spy,backgrounds}
\usetikzlibrary{external}
\usetikzlibrary{patterns}
\usetikzlibrary{shapes,arrows}
\usetikzlibrary{shadows.blur}
\usetikzlibrary{shapes.symbols}
\usetikzlibrary{
	pgfplots.groupplots,
	matrix
}

\newcommand{\black}[1]{{\textcolor[rgb]{0, 0, 0}{#1}}}

\theoremstyle{definition}

\ifCLASSINFOpdf
\else
\fi
\IEEEaftertitletext{\vspace{-1\baselineskip}}

\begin{document}
\title{Efficient Decoders for Short Block Length Codes\\ in 6G URLLC}

\author{Chentao~Yue, Vera~Miloslavskaya, Mahyar~Shirvanimoghaddam, Branka~Vucetic, Yonghui~Li

\thanks{Chentao Yue (corresponding author), Vera Miloslavskaya, Mahyar Shirvanimoghaddam, Branka Vucetic, and Yonghui Li are with the School of Electrical and Information Engineering, the University of Sydney, NSW 2006, Australia.\\
The work of Yonghui Li was supported by ARC under Grant DP190101988 and DP210103410.
}}

\markboth{}%
{Shell \MakeLowercase{\textit{et al.}}: Bare Demo of IEEEtran.cls for IEEE Journals}
%



\maketitle

\begin{abstract}
This paper reviews the potential channel decoding techniques for ultra-reliable low-latency communications (URLLC). URLLC is renowned for its stringent requirements including ultra-reliability, low end-to-end transmission latency, and packet-size flexibility. These requirements exacerbate the difficulty of the physical-layer design, particularly for the channel coding and decoding schemes. To satisfy the requirements of URLLC, decoders must exhibit superior error-rate performance \black{and} low decoding complexity. \black{Also, it is desired that decoders be universal} to accommodate various coding schemes. This paper provides a comprehensive review and comparison of different candidate decoding techniques for URLLC in terms of their error-rate performance and computational complexity for structured and random short codes. We further make recommendations of the decoder selections and suggest several potential research directions.

\end{abstract}


%
\IEEEpeerreviewmaketitle

\vspace{-0.3em} 
\section{Introduction}
\vspace{-0.3em} 

The third generation partnership project (3GPP) has defined ultra-reliable low-latency communications (URLLC) as a critical communication scenario for beyond 5G and 6G networks.  
URLLC in 6G requires a significantly lower end-to-end latency (\black{25$\mu$s $\sim$ 1ms}) compared to the 5G new radio (NR), and a high level of transmission reliability, requiring a block error rate (BLER) \black{achieving $10^{-5}\sim10^{-7}$} \cite{tataria20216g}. 
URLLC will enable emerging applications, such as tactile internet, augmented reality, and metaverse, as well as mission-critical applications, such as telesurgery and factory automation \cite{Mahyar2019ShortCode1}.

Short block-length codes with strong error-correction capabilities are essential in URLLC to meet the stringent latency and reliability requirements \cite{Mahyar2019ShortCode1}. However, the use of short block-length codes could degrade the transmission reliability. According to the normal approximation (NA) bound for the finite block-length regime \cite{erseghe2016coding}, the theoretical maximum ratio of information bits to coded bits that can be \black{transmitted with a given error probability} over a noisy channel significantly drops as the block-length decreases. As a result, codes with short block-lengths frequently exhibit degraded BLER performance, when compared to codes with large blocks and the same code rate. The design of coding schemes for URLLC, therefore, presents a challenging trade-off between block-length and reliability.

Several short block-length codes have been thoroughly reviewed as candidates for URLLC in \cite{Mahyar2019ShortCode1} 
Among the candidates, eBCH codes, cyclic-redundancy-check-aided polar (CRC-polar) codes \black{and polarization-adjusted convolutional (PAC) codes} have shown the ability to approach the NA bound (with gaps less then 0.5 dB at BLER $=10^{-4}$) \cite{Mahyar2019ShortCode1}. 
Turbo codes are not preferred in URLLC due to their severe performance degradation in short block-length regime \cite{Mahyar2019ShortCode1}. LDPC codes have been adopted for data channels of 5G NR; however, they have a gap larger than 1 dB to NA at short block-lengths \cite{Mahyar2019ShortCode1}.

Not only does URLLC require superior codes, but also novel low-complexity decoders. Novel decoding algorithms and their efficient implementations are required to achieve the near maximum-likelihood (ML) performance of the codes while keeping the decoding complexity as low as possible. Generally, codes approaching the NA bound lack efficient decoders. Even though CRC-polar codes can be decoded with a near ML performance using the successive cancellation list (SCL) algorithm \cite{tal2015list}, the decoder requires a large list size leading to a high computational complexity.

Besides, URLLC necessitates the bit-level granularity of the block-lengths and the coding rates to accommodate scenarios with varying latency and bandwidth constraints \cite{Mahyar2019ShortCode1}. Universal decoders can meet the requirement of bit-level granularity for any rate-matching scheme due to their capability to decode any binary linear codes. This allows to consider the best known linear codes of flexible block-lengths and rates for URLLC \cite{Grasslcodetables}, as well as the near optimal design of rate-compatible (RC) codes for incremental-redundancy  hybrid automatic repeat request (IR-HARQ) schemes. Advanced ordered-statistics decoding (OSD) and Guessing Random Additive Noise Decoding\footnote{A. Valembois and M. Fossorier suggested a similar decoding method in 2001.} (GRAND) have been recently proposed as candidate universal decoders for URLLC \cite{Fossorier1995OSD,yue2021probability, duffy2019capacity}. Both are capable of decoding any linear block code. 

In this article, we investigate the OSD, GRAND, SCL and their variations as the main candidate decoding schemes for URLLC. We apply them to decode short block-length codes with high error-correction capability, including short eBCH, CRC-polar\black{, PAC and} random codes of lengths 64 and 128. The decoders are evaluated and compared in terms of their complexity and BLER performance via simulations. We show that universal decoders (e.g., OSD and GRAND) are efficient for near-ML decoding not only of structured eBCH codes, but also of unstructured random codes with the same level of complexity. 
\black{According to the simulation results, the SCL based sequential decoder 
is the best decoder in terms of the performance-complexity trade-off among all compared schemes, whereas universal decoders still need further improvements to work efficiently.} Finally, we identify several research directions for optimizing the decoders for URLLC.

\vspace{-0.3em} 
\section{Performance Metrics and Benchmarks}
 \vspace{-0.3em} 
We outline 
the optimality \black{and} complexity 
as the key performance metrics for decoder design for URLLC\black{, and discuss the desirability of the decoding universality.}

\subsubsection{Optimality} Optimality is defined as the ability of decoders to achieve the code ML performance. Theoretically, the ML decoding requires considering all codewords in the codebook and selecting the codeword with the highest likelihood of matching the received noisy signal. URLLC applications should use short codes with strong error-correction capability to meet requirements of ultra-reliable transmission. Therefore, the employed decoders must be (near) optimal to fully exploit the error-correction capability of the selected short codes. As an example, the half-rate length-128 LDPC code shows a 1.75dB gap to the NA bound under the belief propagation (BP) decoding, while this gap can be shortened to 0.5 dB under near-ML OSD decoding \cite{Mahyar2019ShortCode1}.

\subsubsection{Complexity} Low complexity of decoding is crucial to fulfill the low-latency requirement. In 6G URLLC, the end-to-end latency, defined as the time to successfully deliver a data block from the transmitter to the receiver via the radio interface, is required to be less than 1 ms \cite{tataria20216g}. The end-to-end latency consists of time-to-transmit latency, propagation delay, processing latency, multiple access latency, and re-transmission time. The processing latency includes the latency introduced by channel estimation, detection, decoding, etc. The decoding latency dominates the processing latency. Thus, in URLLC, to conserve the overall budget of latency, the complexity of the decoder needs to be as low as possible.

\subsubsection{Universality} 
The universality of decoders refers to the ability to decode any linear block codes. 
\black{The desirability of decoding universality stems from} (i) the bit-level granularity of the block-length and coding rate desired in URLLC, and (ii) the fact that the best known codes\footnote{These codes have the highest minimum Hamming distances among the state-of-the-art linear codes, and therefore they have a superior performance.} of different lengths and rates have different structures \cite{Grasslcodetables}. Conventionally, the design of codes with flexible parameters (as well as RC codes) relies on rate-matching techniques such as puncturing, shortening and repetition, to vary the code length and rate of well-structured codes. Such highly-structured design results in suboptimal error-correction capabilities of the resulting codes and unnecessarily high decoding complexity. For example, when shortened polar codes from 5G NR are processed by the SCL, the decoding is actually performed on a longer mother code, leading to unnecessary decoding overhead. Instead, universal decoders allow the use of the best known short codes with flexible parameters, as well as the optimal short rate-compatible codes for IR-HARQ. 

\vspace{-0.3em} 
\section{Candidate Universal Decoding Algorithms for URLLC}
 \vspace{-0.3em} 
We briefly introduce several decoders that can be potentially applied in URLLC. In this paper, we use $\mathcal{C}(n,k)$ to denote a binary linear block code with block-length $n$, information block-length $k$, and the code rate $k/n$. We consider \black{AWGN channel with QPSK modulation. The energy per transmitted symbol is $E_s=1$. For a modulated codeword $\mathbf{x}$ of $\mathcal{C}(n,k)$, the received vector is $\mathbf{y} = \mathbf{x} + \mathbf{w}$, where elements of the noise vector $\mathbf{w}$ follow complex Gaussian distribution with zero mean and variance equal to the one-sided power spectral density $N_0$.} 
\black{The SNR in dB is accordingly defined as $10\log_{10}(E_s/N_0)$ and for clarity denoted as $E_s/N_0\mathrm{[dB]}$.}



\vspace{-0.3em} 
\subsection{Ordered-Statistics Decoding (OSD)} \label{sec::candidate::OSD}
 \vspace{-0.3em} 
OSD, first proposed in \cite{Fossorier1995OSD}, is a near-ML universal decoding algorithm. OSD begins each decoding operation by permuting the received symbols in the descending order of magnitudes of their \black{reliabilities}. 
Then, the same permutation is applied to the columns of the generator matrix of $\mathcal{C}(n,k)$, \black{and the resulting matrix is transformed to the systematic form using Gaussian elimination (GE).} 
\black{After that}, OSD flips these $k$ most reliable bits by XORing them with a length-$k$ 
test error pattern (TEP).   \black{Finally}, the $k$ most reliable bits (flipped by TEP) are re-encoded to recover the remaining $n-k$ bits with lower reliabilities.

OSD processes a specific group of TEPs during each decoding round\black{;} TEPs are processed in the ascending order of the Hamming weight from zero to a specified maximum, 
 referred to as the decoding order. As verified in \cite{Fossorier1995OSD}, an OSD decoder of order $m \!=\! \lceil d_{\mathrm{H}}/4\!-\!1\rceil$ approaches the ML decoding performance for codes with the minimum Hamming distance $d_{\mathrm{H}}$. Therefore, OSD could be a turn-key decoding solution; by setting the decoding order to $m$, OSD can always return the ML decoding results.

OSD has a significantly reduced complexity compared to the brute-force ML decoding \cite{Fossorier1995OSD}, because transmission errors are rare among highly reliable bits, necessitating a smaller number of TEPs. Recent years have seen a significant progress towards improvements of OSD, e.g., the probability-based OSD (PB-OSD) \cite{yue2021probability}, and Box-and-Match algorithm (BMA) \cite{FossorierBoxandMatch}. 
\black{They reduced the decoding complexity by eliminating useless TEPs.}

 \begin{figure*} [t]
     \centering
    \includegraphics[scale=1.01]{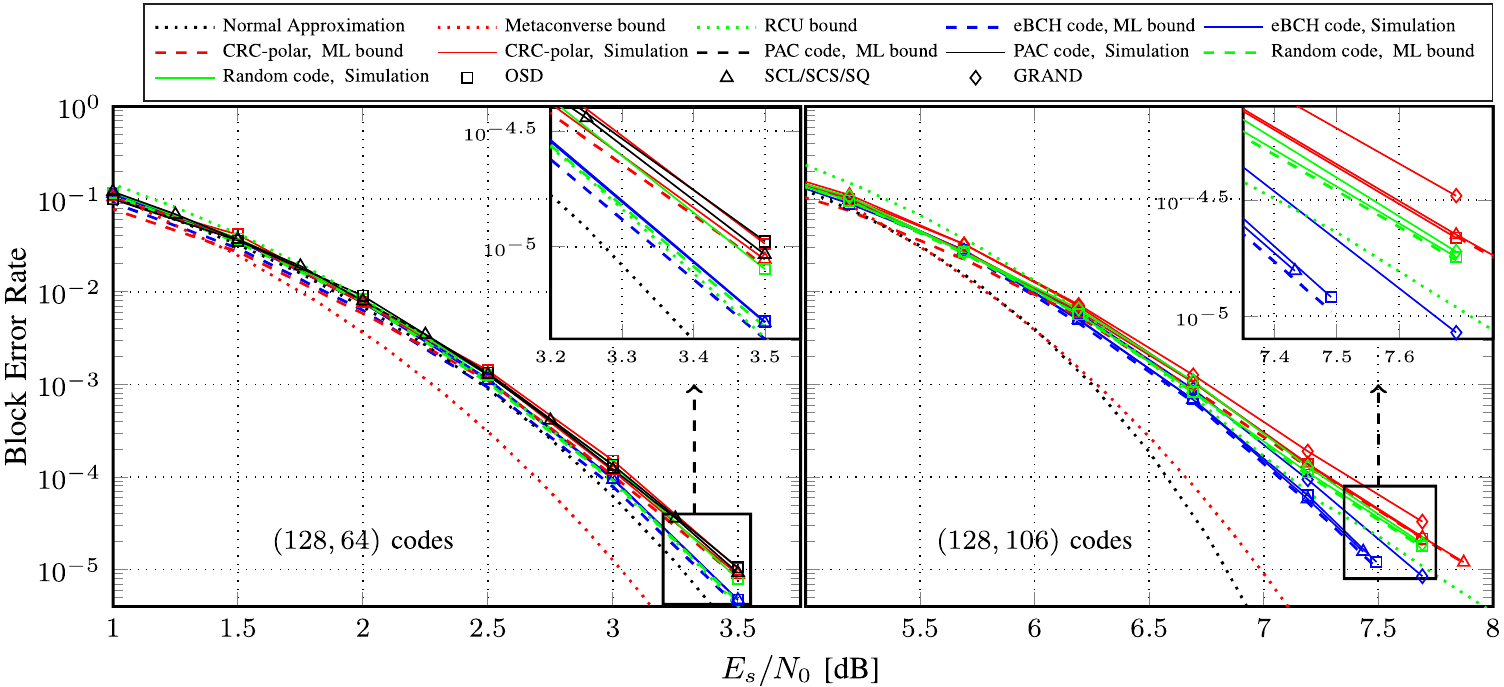}
	\vspace{-0.3em}
    \caption{The BLER performance of eBCH, CRC-11 polar, \black{PAC,} and random codes of length $n=128$, decoded by OSD, GRAND, SCL, SCS, \black{SQ}. 
    }
	\vspace{-1em}
	\label{Fig::length-128::BLER}
        
\end{figure*}

\vspace{-0.3em} 
\subsection{Guessing Random Additive Noise Decoding (GRAND)}
 \vspace{-0.3em} 
The recently proposed GRAND is a universal decoder capable of achieving the ML performance \cite{duffy2019capacity}. Taking the binary symmetric channel (BSC) as an example, GRAND guesses a possible noise sequence $\mathbf{w}'$ to retrieve $\mathbf{c}' = \mathbf{y} \oplus \mathbf{w}'$, and $\mathbf{c}'$ is a valid estimate if  \black{$\mathbf{c}'\in$} $\mathcal{C}(n,k)$, which 
\black{is} checked by computing the syndrome. GRAND checks a group of noise sequences, each check is referred to as a noise query. Therefore, if GRAND performs noise queries in the descending order of the likelihoods of noise sequences, \black{then} the first discovered valid codeword estimate is exactly the ML estimate \cite{duffy2019capacity}. 
For any memoryless channel, the first codeword discovered by GRAND is the ML estimate when 
noise sequences are queried in the strict descending order of their \black{exact} likelihoods.

Intuitively, GRAND is efficient when SNR or the coding rate $k/n$ is high. High SNRs ensure the low Hamming weight of the true noise sequence, while high coding rates increase the probability of passing codebook membership checks. 
\black{As verified in \cite{duffy2019capacity}, GRAND satisfies the ML criterion} with $2^{n\min(H_{\frac{1}{2}},1 - \frac{k}{n})}$ noise queries for a binary code in BSC, where $H_{\frac{1}{2}}$ is \black{the Rényi entropy 
that decreases as the SNR increases. 
The complexity of GRAND can be reduced by limiting the number of noise queries allowed. 
That is, performing decoding until a specified maximum number of noise queries is reached 
and returning a decoding failure if no codeword is found}. \black{However}, GRAND requires querying noise sequences sorted by descending likelihood, which is difficult to perform on the fly. In \cite{duffy2021ordered1} (ordered reliability GRAND), noise sequences are generated offline by ascending their logistic weights \cite[Section II]{duffy2021ordered1}, resulting in an acceptable BLER performance and computational complexity.

\vspace{-0.3em} 
\subsection{Successive Cancellation List \black{(SCL)\black{, Successive Cancellation Stack (SCS)} and Sequential (SQ)} \black{Decoding}}
 \vspace{-0.3em} 





SCL \cite{tal2015list}\black{, SCS \cite{niu2012stack}} 
\black{and SQ \cite{miloslavskaya2014sequential,Trifonov2018ASF}} 
significantly improve the BLER performance compared to the successive cancellation (SC) decoding, which is native for polar codes. SC performs a depth-first search over the code tree. At each level of the tree, a decision on the information bit value is made based on the recursively computed probabilities of `0' and `1' branches. This greedy approach ensures a low complexity to SC decoding but sacrifices its BLER performance. 
SCL keeps up to $L$ paths at each level of the tree, and then outputs the most probable path among fully constructed paths as the final result. SCL can achieve the BLER performance of ML decoding at the expense of increased complexity by using a sufficiently large $L$. 
\black{In contrast with SCL, the sequential decoders 
extend a single path at a time. This reduces  exploration of fruitless  low-probability paths. 
} 
\black{In the SCL based \black{SCS and} SQ,} parameter $L$ limits the number of \black{stored} paths of the same length. 
\black{A very low space complexity is achieved by the Fano based sequential decoder 
that moves not only forward but also backward, i.e., it can not only extend a path but also shorten it. However, the backward movement induces additional time complexity.} 

SC, SCL\black{, SCS} and \black{SQ} decoding techniques can be applied not only to polar codes, but also to any binary linear block code \black{with length equal to a power of two. 
} 
However, the \black{decoding} performance and complexity substantially depend on the code structure. Due to their prohibitively high computational complexity of near ML decoding of unstructured random codes, SCL\black{, SCS} and \black{SQ} cannot be regarded as universal decoders.

\vspace{-0.3em} 
\section{Comparison of Decoders for URLLC} \label{sec::Cmp}
 \vspace{-0.3em} 

In this section, we compare the performance and complexity of various decoders for short eBCH, CRC-polar, \black{PAC,} and random codes. 

\vspace{-0.3em} 
\subsection{BLER Performance}
 \vspace{-0.3em}

 \begin{figure*}  [t]
     \centering
    \includegraphics[scale=1.01]{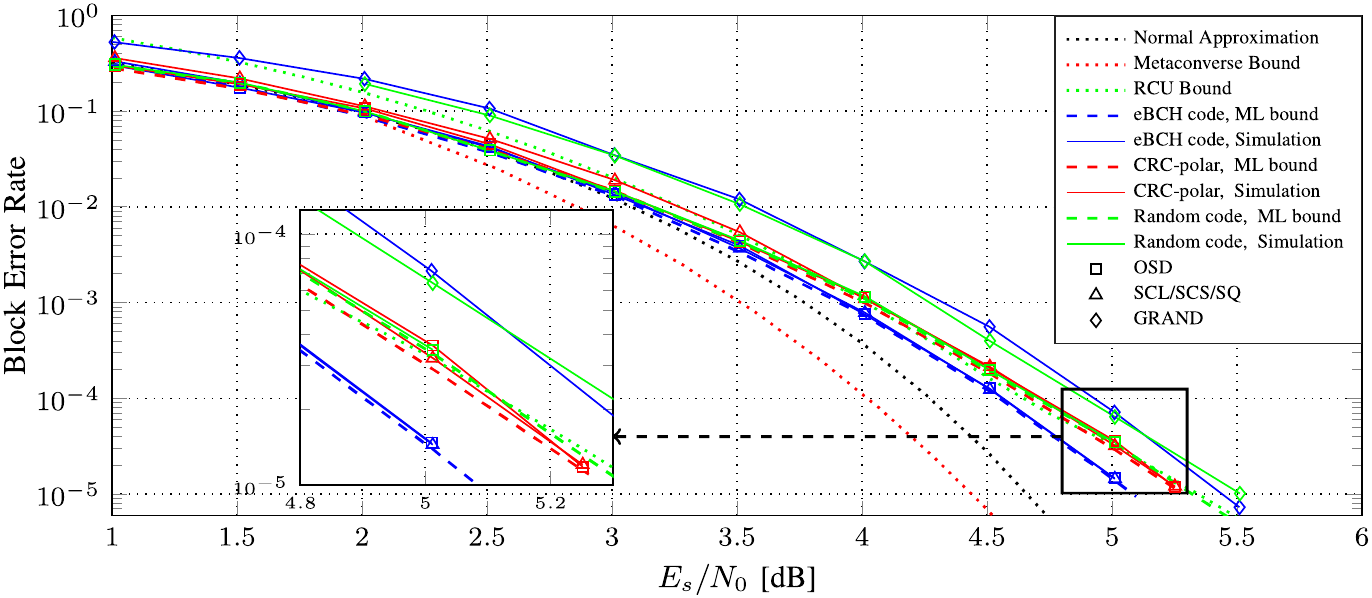}
	\vspace{-0.5em}
    \caption{The BLER performance of $(64,36)$ eBCH, CRC-11 polar and random codes, decoded by OSD, GRAND, SCL, SCS, \black{SQ}. 
    }
    \vspace{-1em}
	\label{Fig::length-64::BLER}
        
\end{figure*}    

    We examine the BLER performance of various decoders for length-128 and length-64 codes \footnote{\black{The considered code parameters fall within the uplink and downlink control channel scenarios of 5G NR, which have the minimum information length of 12 bits and the maximum information lengths of 1706 and 140 bits, respectively. 6G is in the conceptualization stage, and 6G URLLC information lengths have not been yet defined.
    }}. \black{Precisely}, we implement the original OSD algorithm with the optimal decoding order $m = \lceil d_{\mathrm{H}}/4-1\rceil$ \cite{Fossorier1995OSD}. The GRAND decoder is implemented as described in \cite{duffy2021ordered1}, where noise queries are sorted offline according to logistic weights of the corresponding noise sequences. The maximum number of noise queries is set to $10^9$. SCL\black{, SCS} and \black{SQ} are implemented according to \cite{tal2015list}\black{, \cite{niu2012stack}} and \black{\cite{miloslavskaya2014sequential,Trifonov2018ASF}}, respectively. The SCL/\black{SCS/}\black{SQ} list size is at most $L = \black{32768}$, which is sufficient to obtain the near ML performance. By checking if the observed decoding errors are ML errors, we obtain a BLER lower bound under the ML decoding. This bound is further referred to as the ML bound. \black{Normal approximation \cite{erseghe2016coding}, metaconverse \cite{erseghe2016coding}, and random coding union (RCU) \cite{font2018saddlepoint1} bounds are included as performance benchmarks of finite blocklength code} \footnote{\black{RCU and metaconverse bounds are the upper and lower bounds on the best error probability achieved by $(n,k)$ codes, respectively (see \cite[Theorem 1-2]{erseghe2016coding}).}}. 
    
    \subsubsection{Length-128 codes}

    Figure \ref{Fig::length-128::BLER} illustrates the performance of various decoders for different $(128,64)$ and $(128,106)$ codes. CRC-11 is applied to polar codes as defined in the 5G NR. \black{The $(128,64)$ PAC code is constructed as proposed by E. Arıkan in 2019. 
    } The BLER performance results of SCL\black{, SCS} and \black{SQ} are combined because they perform \black{almost} identically in terms of BLER with the same list size \black{for the considered short-length codes}. GRAND is used only for high-rate $(128,106)$ codes, as its complexity for $(128,64)$ codes is prohibitively high. As shown by Fig. \ref{Fig::length-128::BLER}, the $(128,64)$ eBCH\black{, CRC-polar and PAC codes} exhibit very similar ML BLER performance, 
    \black{which} can be approached with the OSD and SCL/\black{SCS/}\black{SQ}. It is worth noting that the $(128,64)$ random code also performs closely to the eBCH code under OSD.
    
    On the other hand, the performance of different $(128,106)$ codes diverges at high SNRs; eBCH outperforms both CRC-polar and the random code
    , while the random code performs similarly to the CRC-polar. At high SNRs, OSD closely approaches the ML performance for all considered codes, whereas GRAND exhibits a 0.1 dB performance gap relative to the ML bounds of the eBCH and CRC-polar codes. We highlight that the performance gap of GRAND is caused by the suboptimal ordering of noise quires. Specifically, when logistic weights are used as likelihoods, noise sequences cannot be queried in the strict descending order of their true probabilities. 
    In contrast, when decoding the $(128,106)$ random code, the performance of GRAND is virtually identical to that of OSD and ML bound. Note that SCL/\black{SCS/}\black{SQ} is not compatible with random codes due to prohibitively high computational complexity. 
    
    \subsubsection{Length-64 codes}
    
    Figure \ref{Fig::length-64::BLER} illustrates the BLER performance of different $(64,36)$ codes, decoded by OSD, SCL/\black{SCS/}\black{SQ}, and GRAND. As shown, eBCH has the best ML BLER performance compared to its counterparts, and its ML bound is reached by OSD and SCL/\black{SCS/}\black{SQ}. For eBCH, GRAND has a gap of around 0.4 dB to the ML bound at high SNRs. 
    The CRC-polar and random code have a similar performance for all considered decoders; however, they both perform worse than eBCH, losing 0.25 dB of coding gain at BLER = $10^{-5}$. Note that SCL/\black{SCS/}\black{SQ} are designed for polar codes, but they are also capable of decoding eBCH codes with near-optimal performance.

     \begin{figure*}  
     \centering
    \includegraphics[scale=1.01]{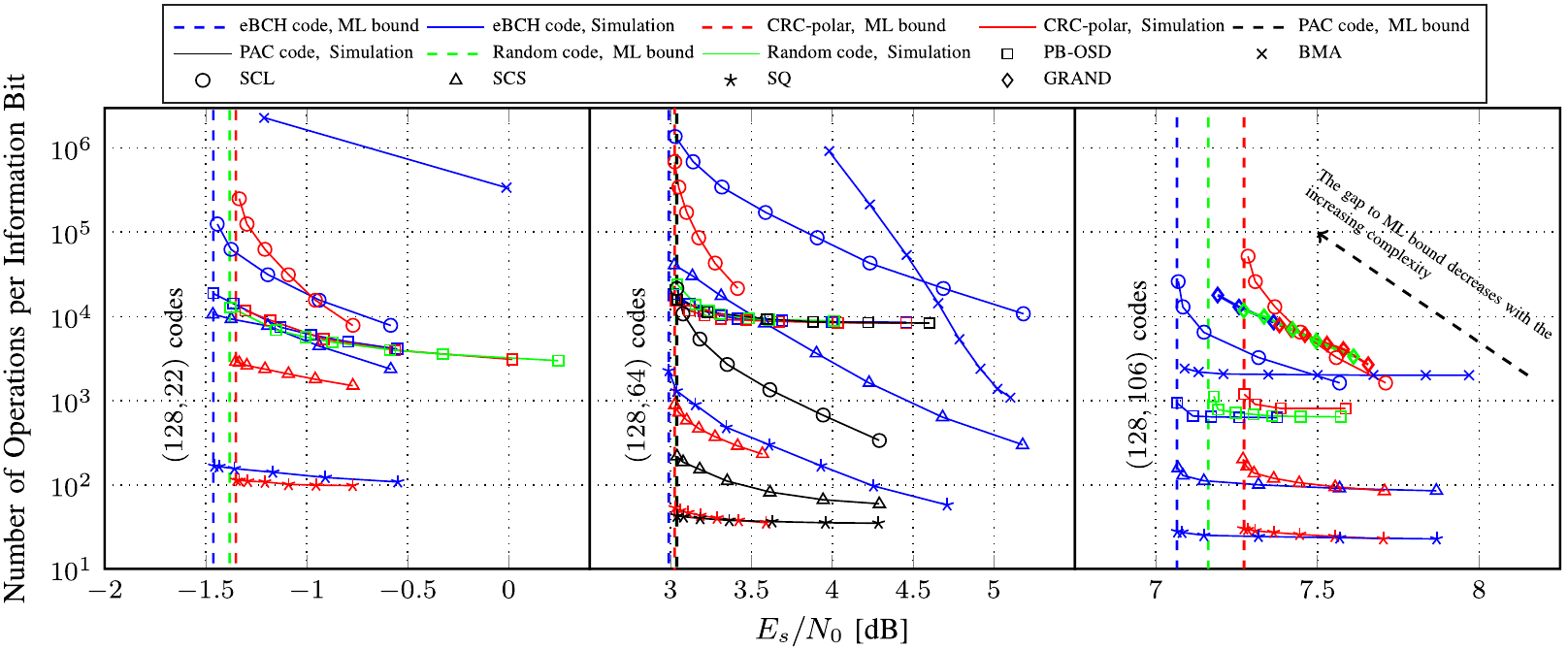}
	\vspace{-0.5em}
    \caption{Complexity versus the SNR required to achieve BLER = $10^{-4}$ in decoding codes of length 128 using different decoders. 
    }
    \vspace{-1em}
    \label{Fig::length-128::GaptoML}     
    \end{figure*}

\vspace{-0.3em} 
\subsection{Performance-Complexity Trade-offs} \label{Sec::Complexity}
 \vspace{-0.3em} 

    We compare the average decoding complexity in terms of the number of operations per information bit. The number of operations is defined as the total number of additions, multiplications, and comparisons. The trade-offs between the BLER performance and complexity \black{are} controlled by adjusting a few decoder parameters to improve the BLER performance at the expense of increased complexity. 
    
    For the comparison, we implemented the recently proposed PB-OSD \cite{yue2021probability} and the well-known BMA \cite{FossorierBoxandMatch} as variations of the original OSD decoder. Their complexity can be configured by restricting the number of TEPs; that is, setting a threshold success decoding probability for PB-OSD \cite[Eq. (6)]{yue2021probability} and the cardinality of the control band (CB) for BMA \cite[Section III-A]{FossorierBoxandMatch}. GRAND is implemented with a range of limits on the maximum number of noise queries. A small number of noise queries leads to a reduced decoding complexity and an increased BLER.  Each data point in Figs. \ref{Fig::length-128::GaptoML}-\ref{Fig::length-64::GaptoML} is produced by running simulations until 1000 decoding errors.

    \subsubsection{Length-128 codes}
    
    In Fig. \ref{Fig::length-128::GaptoML}, we show SNRs required to achieve the BLER of $10^{-4}$ for length-128 codes using various decoders. Each decoder is characterized by a curve obtained by varying its parameters, e.g., list size $L$ for SCL/\black{SCS/}\black{SQ}, threshold probability for PB-OSD, CB cardinality for BMA, and query limit for GRAND. As a universal decoder, PB-OSD has nearly unvaried complexity for different codes of the same rate, which is also the case for the universal GRAND. In contrast, SCL\black{, SCS} and \black{SQ} heavily rely on the code structure. 
    
    At a fixed SNR, GRAND has the highest complexity in terms of operations per information bit for $(128,106)$ eBCH code, while \black{SQ} has the lowest complexity among all competitors. However, GRAND \black{and OSD} can fully exploit the advantage of parallelism in processing noise queries \black{and TEP, respectively}; if multiple parallel noise processors are applied, the \black{decoding} time 
    will be significantly reduced. 
    In contrast, parallelization of SCL\black{, SCS} and \black{SQ} is much more sophisticated due to the sequential nature of the SC decoding. \black{Due to 
     various parallelization opportunities, the decoding complexity depicted in Fig. \ref{Fig::length-128::GaptoML} (as well as in Fig. \ref{Fig::length-64::GaptoML}) may not necessarily represent the decoding latency.} The detailed description of parallel decoders is beyond the scope of this paper.

	\subsubsection{Length-64 codes}
	
\begin{figure*}  [t]
         \centering
        \includegraphics[scale=1.01]{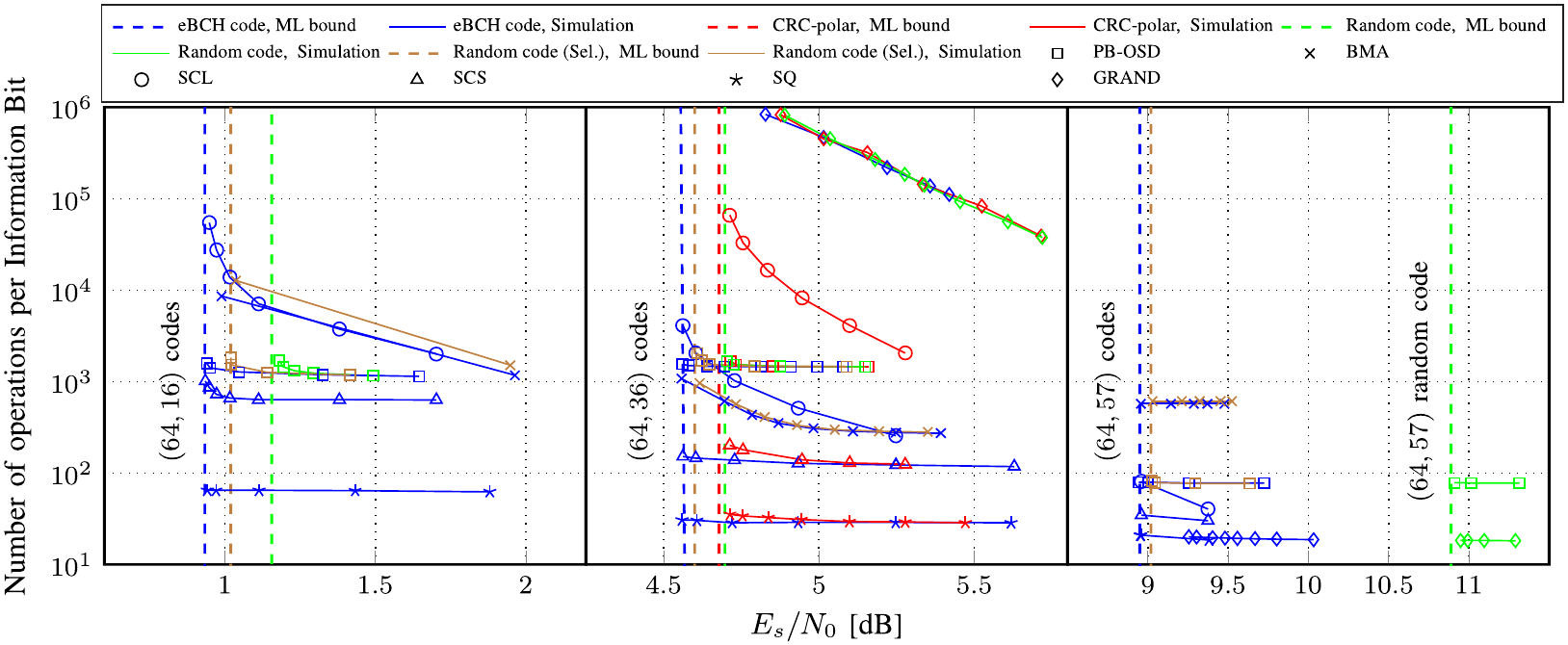}
     	\vspace{-0.5em}
            \caption{Complexity versus the SNR required to achieve BLER = $10^{-4}$ in decoding codes of length 64 using different decoders. 
        }   
        \vspace{-0.9em}
    \label{Fig::length-64::GaptoML}
\end{figure*}

	Fig. \ref{Fig::length-64::GaptoML} compares the SNRs required to achieve the BLER of $10^{-4}$ for length-64 codes under different decoders. Only the $(64,36)$ CRC-polar code is simulated, since the $(64,16)$ and $(64,57)$ \black{polar codes with CRC-11} are not defined in the 5G NR. \black{The $(64,57)$ eBCH code is equivalent to the Reed-Muller code, and therefore pure polar code. $(64,16)$ and $(64,36)$ pure polar codes} are not considered due to their inferior BLER performance \cite{Mahyar2019ShortCode1}. 
	
	As shown in Fig. \ref{Fig::length-64::GaptoML}, the $(64,57)$ random code suffers from a significant performance degradation compared to the eBCH code. \black{We improve the performance of a short random code by sequentially selecting the random generator/check matrix columns using a greedy approach.} Specifically, each column \black{is} selected as follows. Given a number of randomly generated candidate columns, their effects on the BLER are estimated by using the union bound and then the best candidate is appended to the generator/check matrix until all columns are selected. 
	As shown by Fig. \ref{Fig::length-64::GaptoML}, such a selected random code can perform similarly to the eBCH code at all rates under PB-OSD, and significantly outperforms the pure random codes.

	\black{It can also be seen that SQ is remarkably efficient in decoding eBCH codes. 
    PB-OSD and BMA can be used for decoding $(64,16)$ and $(64,36)$ random codes.} 
    Moreover, it can be seen that the complexity of all universal decoders, including BMA, PB-OSD, and GRAND, is independent of the code structure and is defined by the block-length $n$ and information length $k$.

	\vspace{-0.3em} 
    \subsection{Other Considerations for Decoding Complexity}	
  \vspace{-0.3em} 
    
    \subsubsection{Complexity of near ML Decoders Operating at Different SNRs}
\begin{figure}  [t]
     \centering
        \includegraphics[scale=1.0]{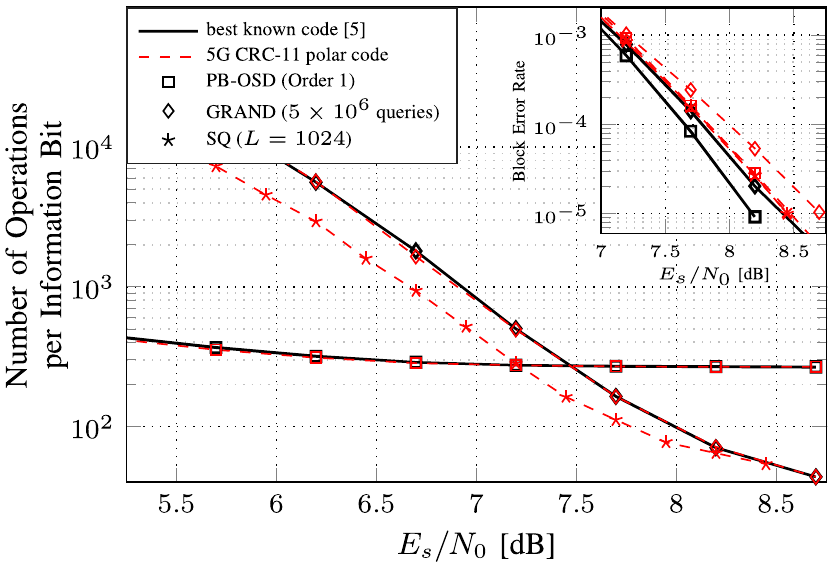}
	\vspace{-0.5em}
    \caption{BLER and complexity of various decoders operating at different SNRs for the $(82,68)$ 5G CRC-11 polar code and the best known code from \cite{Grasslcodetables}. PB-OSD \black{and SQ} are configured to have the near ML performance of BLER. GRAND has a gap to the ML performance due to the suboptimal noise query ordering.
    }
    \vspace{-0em}
	\label{Fig::complexity::SNR}
        
	\end{figure}

    In Fig. \ref{Fig::complexity::SNR}, we demonstrate the complexity of decoders operating at different SNRs for the $(82,68)$ CRC-11 5G shortened polar code and the best known code from \cite{Grasslcodetables}. As shown, decoders typically have a lower complexity at higher SNRs. 
    \black{At low SNRs, PB-OSD has the lowest complexity of all decoders. 
    GRAND and SQ have the highest complexity at low SNRs, but their complexity decreases rapidly as the SNR increases. They are more efficient than OSD at extremely high SNRs because they avoid complex steps such as Gaussian elimination.}

    \subsubsection{Worst-Case Complexity}
    
    The worst-case complexity matters in URLLC scenarios. When some received blocks have a low degree of reliability, the decoder may require a much higher level of complexity than average to produce a decoding result. In practical implementations, the worst-case decoding complexity is typically restricted. For example, if a decoder exceeds a certain number of operations (e.g., visited tree nodes of SC, TEPs of OSD, and noise queries of GRAND), then decoding can be terminated. Due to the fact that high decoding complexity is usually associated with received blocks that will be incorrectly decoded, early termination techniques in decoders typically result in a negligible performance degradation. As numerous prior works have proposed limiting the complexity of various decoders, we only briefly describe the worst-case complexity from the theoretical perspective. SCL has a constant complexity, so its worst case complexity is equal to its average complexity \cite{tal2015list}. The worst-case complexity of \black{SCS and }\black{SQ} is \black{similar to} the complexity of SCL. 
    The complexity of PB-OSD and BMA is upper bounded by that of the original OSD \cite{yue2021probability}. Finally, according to \cite{duffy2021guessing}, the worst-case complexity of GRAND is very high unless the code rate is close to 1. \black{According to our simulation results, PB-OSD has the lowest worst-case complexity when decoding the $(82,68)$ 5G CRC-11 polar code, followed by 
    SQ, 
    and finally GRAND. }
    
    \subsubsection{Complexity Scaling with Block-Length}
    
    The complexity scaling with the block-length is another issue of concern in traditional decoder design. For instance, the complexity of SC scales with the block-length $n$ as $O (n\log n)$ owing to the recursive structure of polar codes. Accordingly, the complexity of SCL with list size $L$ scales with $n$ as $O (L n\log n)$. \black{The complexity of \black{SCS and} SQ scales similarly at low SNRs, but at high SNRs it approaches $O (n\log n)$.}  
    Because of Gaussian elimination, the complexity of OSD and its variants increases faster than $O(n\min(k,n-k)^2)$. Complexity of GRAND scales approximately as $O(2^{\min(nH_{\frac{1}{2}},n - k)})$ \cite{duffy2019capacity}, which could increase rapidly as the block-length increases. As a result, universal decoders such as OSD and GRAND are better applicable to short block-length codes. 

\vspace{-0.5em} 
\section{Recommendations and Potential Research Directions}
\vspace{-0.3em} 

To exploit the error-correction capabilities of short block codes and achieve high reliability and low latency, we make several recommendations and identify potential research directions to design decoder for in URLLC.

\vspace{-0.8em} 
\subsection{Rate-Compatible Codes with Near Optimal Performance} \label{sec::Rec::1}
\vspace{-0.4em}

 \begin{figure}  [t]
     \centering
    \includegraphics[scale=1.0]{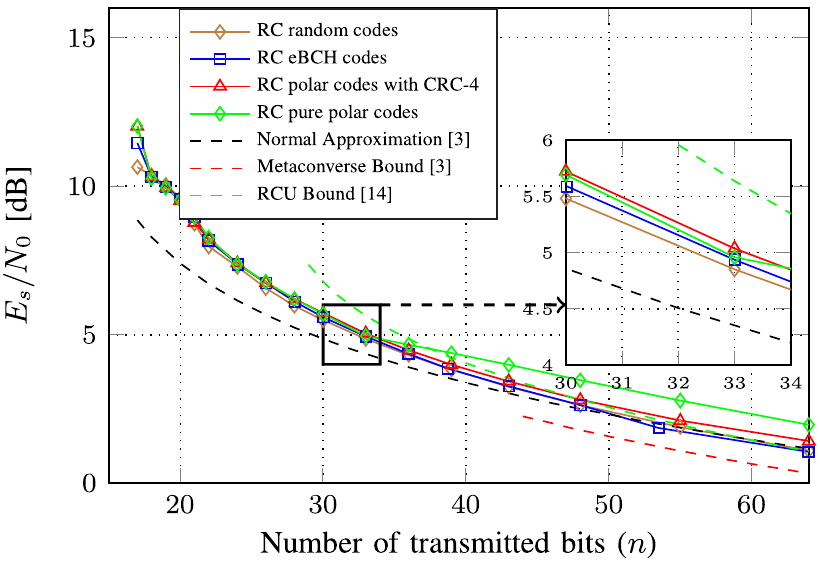}
	\vspace{-0.5em}
    \caption{SNRs required to achieve BLER = $10^{-4}$ for different RC code under the ML decoding. The information block-length is fixed to $k = 16$, and the code block-length varies from $n=17$ to $n=64$.}
    \vspace{-0em}
	\label{Fig::SNRtoNA}
        
\end{figure}

As shown in Section \ref{sec::Cmp}, the universal decoders are capable of decoding unstructured codes (e.g., random codes) with the same level of complexity as structured codes (e.g., eBCH). Therefore, 
one can directly employ codes with the strongest error-correction capability with universal decoders (see tables of the best known codes \cite{Grasslcodetables}).

On the other hand, the  near-optimal RC codes\footnote{Codes are referred to as rate-compatible if they have the same information block length $k$ and their $k\times n$ generator matrices are nested.} for IR-HARQ can be designed for universal decoders.
 Fig. \ref{Fig::SNRtoNA} depicts the SNRs required to achieve BLER of $10^{-4}$ for different rate-compatible codes with $k=16$ and $17\leq n \leq 64$. The RC eBCH, polar, and CRC-4 polar codes are obtained by puncturing their $(64,16)$ mother codes. The RC random codes are constructed by the selection process described in Section \ref{Sec::Complexity}, i.e., the randomly generated columns are sequentially selected and appended to the generator matrix. As shown by Fig. \ref{Fig::SNRtoNA}, such a simple selection process can generate high-performance short RC codes, whose error-correction capability outperforms that of RC eBCH (at half to high rates), RC polar, and RC CRC-polar codes. For universal decoders, we recommend to design near-optimal short RC codes by optimizing the minimum Hamming distance and weight spectrum to satisfy the requirements of the high reliability and bit-level granularity for URLLC.

\vspace{-0.7em} 
\subsection{Enhancement of Universal Decoders}  \label{sec::Rec::2}
\vspace{-0.3em} 

As shown by Figs. \ref{Fig::length-128::GaptoML}-\ref{Fig::length-64::GaptoML}, OSD has a reasonable trade-off between complexity and BLER for low-to-moderate-rate codes with different structures. \black{However, as demonstrated by Fig. \ref{Fig::complexity::SNR}, 
the overhead of Gaussian elimination 
causes OSD to exhibit a ``complexity floor'' even at high SNRs, which 
precludes the application of OSD in scenarios requiring a BLER of less than $10^{-5}$.} Therefore, 
further improvement of OSD, especially to break its complexity floor, is an important research topic for developing a promising universal near ML decoder for URLLC applications. \black{Besides}, a prototype of an advanced OSD decoder, whose efficiency is demonstrated in real-world settings, is desired. Although TEPs in OSD can be processed in parallel, an advanced OSD prototype design that fully exploits the parallelization opportunities is still lacking.

According to Figs. \ref{Fig::length-64::GaptoML}-\ref{Fig::complexity::SNR}, GRAND is only efficient at decoding high-rate short codes at high SNRs. Thus, additional improvements are required to extend benefits of GRAND to low-rate codes. Furthermore, even for high-rate codes, GRAND exhibits a gap to the ML BLER performance when the noise queries are arranged suboptimally. It is unclear how to resolve this problem since GRAND will loose its low-complexity feature if noise queries are optimally determined on the fly. \black{Therefore,} the application of GRAND in URLLC requires a novel efficient algorithm for arranging noise sequences.

\vspace{-0.7em} 
\subsection{Pairs of Codes and Decoders}
\vspace{-0.3em} 

While universal decoders are available, URLLC applications can also allow for the selection of code-decoder pairs. As illustrated by Figs. \ref{Fig::length-128::GaptoML}--\ref{Fig::length-64::GaptoML}, eBCH code has the best coding gain compared to CRC-polar and random codes\black{, while PAC codes lack an explicit construction for an arbitrary code rate. The selected random codes can perform closely to eBCH with a gap smaller than 0.1 dB. CRC-polar codes of length 128 also provide promising BLER performance at low-to-moderate rates. Therefore, 
we recommend them 
with SQ decoding
, while eBCH 
with SQ can be used for high-rates scenarios, as well as very short block-length scenarios.} 
PB-OSD can be used as the decoder at low-to-moderate SNRs according to Fig. \ref{Fig::complexity::SNR}. Moreover, hybrid systems with multiple decoders can be developed to accommodate different codes with the best trade-off between the complexity and BLER. According to different latency and complexity budgets, the hybrid system can adaptively select the optimum encoder and decoder. However, as discussed in Section \ref{sec::Rec::1}, such a complicated hybrid system can be significantly simplified if efficient universal decoders are available.

\vspace{-0.7em} 
\section{Conclusion}
\vspace{-0.3em} 
This paper compared the best available decoders for URLLC, including OSD, BMA, GRAND, SCL, \black{SCS,} and \black{SQ}. Several short block codes with strong error-correction capabilities were considered for evaluating the performance and complexity of these decoders. We showed that efficient universal decoders such as the improved OSD allow to use binary linear codes with near optimal error-correction capability for URLLC
\black{; nevertheless, they need further decoding complexity improvements.} We also observed that length-128 CRC-polars can achieve an 
\black{excellent performance-complexity trade-off when decoded by SQ. For lower block-lengths, 
eBCH with SQ have a low complexity 
at high SNRs
, while PB-OSD can be used 
at low SNRs according to Fig. \ref{Fig::complexity::SNR}}. We also suggested several research directions  for improving the complexity and performance of decoders in URLLC systems.

\vspace{-0.5em} 
	
\bibliography{reference/IEEEabrv,reference/OSDAbrv,reference/SurveyAbrv,reference/ClassicAbrv,reference/MLAbrv,reference/MathAbrv,reference/GrandAbrv,reference/NOMAAbrv,reference/PolarAbrv,reference/Vera,reference/Compressed}
\bibliographystyle{IEEEtran}  
	
\vskip -1.5\baselineskip plus -1fil	

\begin{IEEEbiographynophoto}{Chentao Yue}
(Member, IEEE) received B.Sc. in information engineering from Southeast University, China, in 2017, and the Ph.D. degree from the University of Sydney (USYD), Australia, in 2021. He is currently a postdoctoral research associate at the School of Electrical and Information Engineering (EIE), USYD. 
\end{IEEEbiographynophoto}

\vskip -1.5\baselineskip plus -1fil

\begin{IEEEbiographynophoto}{Vera Miloslavskaya}
received B.Sc., M.Sc. and PhD degrees from Peter the Great St. Petersburg Polytechnic University (SPbPU) in 2010, 2012 and 2015, respectively. 
She is currently a Postdoctoral Research Associate at the EIE, USYD.   
\end{IEEEbiographynophoto}

\vskip -1.5\baselineskip plus -1fil

\begin{IEEEbiographynophoto}{Mahyar Shirvanimoghaddam}
(Senior Member, IEEE) received the B.Sc. degree from The University of Tehran, Iran, in 2008, the M.Sc. degree from Sharif University of Technology, Iran, in 2010, and the Ph.D. degree from The University of Sydney, Australia, in 2015. He is currently a Senior Lecturer at the EIE, USYD. 
\end{IEEEbiographynophoto}

\vskip -1.5\baselineskip plus -1fil

\begin{IEEEbiographynophoto}{Branka Vucetic} (Life Fellow, IEEE) is an Australian Laureate Fellow, a Professor of Telecommunications, and Director of the Centre for IoT and Telecommunications at USYD. She is a Fellow of the Australian Academy of Technological Sciences and Engineering and the Australian Academy of Science. 
\end{IEEEbiographynophoto}

\vskip -1.5\baselineskip plus -1fil

\begin{IEEEbiographynophoto}{Yonghui Li} (Fellow, IEEE) received his PhD degree in November 2002 from Beijing University of Aeronautics and Astronautics. He is now a Professor and Director of Wireless Engineering Laboratory at the EIE, USYD. 

\end{IEEEbiographynophoto}

\end{document}